\documentclass[aps,twocolumn,prl,amssymb,floatfix,superscriptaddress]{revtex4}

\usepackage{graphicx}
\usepackage{epstopdf}

\begin{document}
\title{Multidimensional instability and dynamics of spin-avalanches in crystals of nanomagnets}

\author{O. Jukimenko}
\affiliation {Department of Physics, Ume{\aa} University, SE-901\,87
Ume{\aa}, Sweden}

\author{C. M. Dion}
\affiliation {Department of Physics, Ume{\aa} University, SE-901\,87
Ume{\aa}, Sweden}

\author{M. Marklund}
\affiliation {Department of Physics, Ume{\aa} University, SE-901\,87
Ume{\aa}, Sweden}
\affiliation{Department of Applied Physics, Chalmers University of Technology, SE-412 96 G\"{o}teborg, Sweden}

\author{V. Bychkov}
\affiliation {Department of Physics, Ume{\aa} University, SE-901\,87
Ume{\aa}, Sweden}

\begin{abstract}
%The switching of a material's magnetic properties on the nanolevel in terms of front propagation has recently gained a large interest.
%However, the theoretical underpinning of such switching is still not well understood.
We obtain a fundamental instability of the magnetization-switching fronts in super-paramagnetic and
ferromagnetic
materials such as crystals of nanomagnets,
ferromagnetic nanowires, and systems of quantum dots with large
spin. We develop the instability theory for both linear and nonlinear
stages. By using numerical simulations we investigate the
instability properties focusing on spin avalanches in crystals of
nanomagnets. The instability distorts spontaneously the fronts and leads to a complex multidimensional front dynamics.  We
show that the instability has a universal physical nature, with a deep
relationship to a wide variety of physical systems, such as the
Darrieus-Landau instability of deflagration fronts in combustion,
inertial confinement fusion and thermonuclear supernovae, and the
instability of doping fronts in organic semiconductors.
%Our results are discussed in the context of recent and future experiments.
\end{abstract}

\maketitle

Advanced magnetic materials with super-para\-mag\-netic and ferromagnetic properties, such as molecular (nano-) magnets, ferromagnetic nanowires and quantum dots with spins larger than $1/2$, are the focus of  active research due to their promising applications to spintronics and quantum data storage \cite{Bogani-Wernsdorfer-2008,Nakatani-NatMat-2003,Beach-NatMat-2005,Misiorny-NatPhys-2013}.
In contrast to classical magnetic dipoles, nanomagnets may keep their spin orientation unchanged in altering magnetic fields \cite{Bogani-Wernsdorfer-2008,Friedman-PRL-96,Thomas-Nature-96}.
Spontaneous transition of a nanomagnet from the metastable state (against the field) to the ground state (along the field) is hindered by
the magnetic anisotropy.
%a considerable potential barrier.
%At sufficiently low temperatures and for the magnetic field pointing along the easy ($z$-) axis, a nanomagnet is in the ground state with the spin directed along the external field (e.g.  $S_{z} = - 10$ for $\rm{Mn}_{12} $-acetate).   Upon flipping the magnetic field direction, the former ground state $S_{z} = - 10$ becomes metastable due to the Zeeman energy increase as illustrated in Fig. 1.  At the same time the nanomagnet may not go over spontaneously to the new ground state ($S_{z} = 10$) because of the potential barrier $E_{a}$ separating the states.
In crystals of nanomagnets, the transition
%from the metastable to the ground state
may be induced  by Zeeman energy release in a spin avalanche,   spreading in the form of a magnetic deflagration front (due to thermal conduction) \cite{Suzuki-05,Hernandez-PRL-05,Garanin-Chudnovsky-2007,McHugh-09,Modestov-2011,Dion-2013,Subedi-2013}
or a magnetic detonation front (due to shock waves) \cite{Dion-2013,Decelle-09,Modestov-det}.
%Within the front, the transition  is facilitated by heating due to Zeeman energy release, which spreads in the crystal because of thermal conduction (deflagration) or shock waves (detonation).
Then, in an external magnetic field, a spin-avalanche front switches
the magnetization of a crystal to the energetically favorable state,
similar to the propagation of a domain wall in ferromagnetic nanowires \cite{Nakatani-NatMat-2003,Beach-NatMat-2005}.

So far, almost all experimental and theoretical studies of spin
avalanches have assumed a simplified planar 1D geometry of the
propagating fronts
\cite{Suzuki-05,Hernandez-PRL-05,Garanin-Chudnovsky-2007,McHugh-09,Modestov-2011,Dion-2013,Subedi-2013}. Only
recently, the possibility of 3D bending of a spin-avalanche front
has been encountered in heavy numerical simulations for the specific
propagation mechanism controlled by the  dipole-dipole interaction
close to the tunneling resonance \cite{Garanin-2013}.
We stress that the propagation mechanism studied in \cite{Garanin-2013} is not related to the temperature gradient across the front, and thus conceptually different from the experimentally observed magnetic deflagration \cite{Suzuki-05,Hernandez-PRL-05,Subedi-2013}.
It has been
suggested in \cite{Garanin-2013} that the 3D bending of the magnetization-switching front is a specific feature of the
dipole-controlled propagation mechanism, and thus may suffer from a
narrow domain of applicability.  Moreover, the very existence of the dipole-controlled propagation mechanism studied in \cite{Garanin-2013} has not yet been confirmed experimentally.
Thus, the issue of multidimensional magnetic deflagration dynamics has remained open.

\begin{figure}
\includegraphics[width=3.3in]{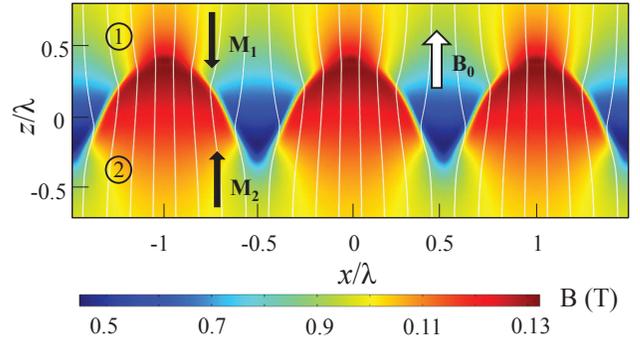}
\caption{\label{fig:mag_field}The magnetic field (shown by colors and field lines) at a 2D curved stationary front propagating at constant speed as obtained using numerical simulations of Eqs. (4), (5) for $\mu_{0}M=0.05\,\textrm{T}$,  $B_{0}=0.1\,\mathrm{T}$ and $\lambda /L_{f}=20$.}
\end{figure}
%\begin{figure}
%\includegraphics[width=3.2in,height=1.6in]{Fig1.eps}
%\caption{The energy levels for the $\rm{Mn}_{12}$-acetate molecule in the external field $B_{z}=1$ T; Calculated similar to \cite{Dion-2013}.}
%\end{figure}
Here we demonstrate that the 3D bending of the magnetization-switching fronts in super-paramagnetic or ferromagnetic materials is a universal physical phenomenon, arising in a common situation when the front propagation speed is controlled by the applied magnetic field.
%We show that the instability is quite general and may arise for different mechanisms of front propagation, e.g., for the thermal conduction mechanism for spin-avalanches in crystals of nanomagnets \cite{Suzuki-05,Hernandez-PRL-05,Garanin-Chudnovsky-2007,McHugh-09,Modestov-2011,Dion-2013,Subedi-2013}, for the non-thermal propagation mechanism considered in Ref. \cite{Garanin-2013}, and for the dissipation-controlled regime of the domain-wall propagation in ferromagnetics \cite{Nakatani-NatMat-2003,Beach-NatMat-2005}.
We find that the instability distorts such fronts and increases their propagation speed. We develop a theory for both the linear and nonlinear stages of the instability, and perform numerical simulations to investigate the instability properties, focusing on spin-avalanche fronts in crystals of nanomagnets. We demonstrate that the instability leads to a complex multidimensional dynamics, with the possibility of stationary cellular structures emerging at the fronts or powerful front acceleration.
Among other conclusions, the present theory explains 3D bending of the
dipole-controlled fronts encountered in the numerical simulations of
Ref. \cite{Garanin-2013} as a particular case. The universal approach to the problem used in our work makes it possible to understand the deep physical relation of the present instability to phenomena from other fields of physics
such as the Darrieus-Landau instability of deflagration fronts in combustion, inertial confinement fusion and thermonuclear supernovae \cite{Law-book,Bychkov-Liberman-2000,Modestov-et-al-2009,DL-supernovae}, as well as  the instability of doping fronts in organic semiconductors \cite{Bychkov-et-al-2011,Bychkov-et-al-2012}.

We consider a generic model of an initially planar front in a magnetically active material.   The front propagation
speed $U_{f}$ depends on the external magnetic field $\textbf{B}$
applied normally to the original front; we set the $z$-axis along the
direction of the external magnetic field. The front modifies the magnetic properties of the
material. In crystals of nanomagnets, for the 2D geometry of Fig. 1, the magnetization vector of a fixed absolute value $M$ switches from $\textbf{M}_{1}\equiv (M_{x1};\,M_{z1})= (0;\,-M)$ ahead of the front (index 1 and label 1 in Fig. 1) to $\textbf{M}_{2}\equiv (M_{x2};\,M_{z2})=(0;\,M)$ behind the front (index 2 and label 2 in Fig. 1).
 Deviations of the magnetization vector from the $\pm z$-axis may be neglected \cite{Dion-2013}.
%and the  magnetic deflagration speed depends on the magnetic field just behind the front, $U_{f}(B_{2})$, see \cite{Suzuki-05,Garanin-Chudnovsky-2007,Dion-2013}.

The physical meaning of the instability may be understood from
Fig.~\ref{fig:mag_field}. The front bending
modifies the magnetic field, with the absolute field value increasing
close to the front humps in agreement with Maxwell's equations. This
increase of the magnetic field is similar to the increase of the electric
field by the convex parts of a conductor (e.g., at a distorted doping
front \cite{Bychkov-et-al-2011,Bychkov-et-al-2012}), or to
modifications of the gas velocity at the humps of a wrinkled flame
front \cite{Law-book}; here the magnetic, electric, and velocity fields
play conceptually the same role.  In turn, the increase of the
magnetic field close to the front humps produces a local increase of
the front speed, thus leading to further unstable growth of the
hump. We demonstrate this effect below by solving the stability
problem for the originally planar magnetic deflagration front
$Z_{f}=U_{f}t$ with the initial magnetic field normal to the front,
$\textbf{B}_{0}=\widehat{\textbf{a}}_{z}B_{0}$. The front position is defined as $z=Z_{f}(\textbf{x},t) $. 
%which works well for the present study, although excludes the general possibility of strongly stirred and multiple fronts.

We consider infinitesimal front perturbations as a superposition of Fourier modes,
$Z_{f}(x,t)= U_{f}t + \widetilde{Z}_{f} ( \textbf{x}, t)$, where
$\widetilde{Z}_{f}(\textbf{x},t)= \sum_k \widetilde{Z}_{k} \exp( i \textbf{k}\cdot \textbf{x} + \sigma t)$ with the perturbation wave number $k=2\pi/\lambda$, the wavelength $\lambda$ and the factor $\sigma$. 
%At the nonlinear instability stage the physical meaning of $\lambda$ is the distance between two neighboring humps at the front.
%Although the linear stability analysis is performed for 2D perturbations, it may be extended to 3D case in a straightforward way be replacing $k x$ by $\textbf{k}\cdot \textbf{x}$ and $k$ by $|\textbf{k}|$.
The purpose of the linear stability problem is to find the dispersion relation $\sigma(k)$; the front is unstable with respect to the bending if $\textrm{Re}(\sigma) >0$ for at least some values of $k$. As we show  below, the factor $\sigma$ is real and positive in this problem, and may be called ``the instability growth rate''. We consider the stability of an infinitely thin front, $kL_{f} \ll 1$, where $L_{f}$ is the front thickness controlled by transport processes, e.g., by  thermal diffusion $\kappa$ in the case of magnetic deflagration, $L_{f} \equiv \kappa / U_{f}$.
Perturbations of the front induce perturbations of the magnetic field both ahead and behind the front, $\textbf{B}=\textbf{B}_{0}+ \sum \widetilde{\textbf{B}}_{k}(z)\exp( i\textbf{k}\cdot \textbf{x} + \sigma t)$, which satisfy Maxwell's equations for a nonconducting medium, $ \nabla \cdot \textbf{B}=0$, $\nabla \times \textbf{H}=0$, $\textbf{B}/\mu_{0}=\textbf{H}+\textbf{M}$,
where $\mu_{0}$ is the vacuum permeability.
  Taking into account the vanishing of the perturbations far away from the perturbed front, at $z\rightarrow \pm \infty$, we solve Maxwell's equations as
$\widetilde{\textbf{B}}_{1,2}(z)\propto \exp( \mp k z)$. We match the solutions using the boundary conditions $\widehat{\textbf{a}}_{n}\cdot \left[\textbf{B}\right]=0$, $\widehat{\textbf{a}}_{n}\times \left[\textbf{H}\right]=0$, where   the normal vector to the perturbed front is $\widehat{\textbf{a}}_{n}=\widehat{\textbf{a}}_{z}-\nabla_{\bot}{Z}_{f}
$, within the linear problem, $\nabla_{\bot}$ corresponds to the transverse variables $\textbf{x}$, and $\left[F\right]\equiv F_{2}-F_{1}$ designates the difference of any value $F$ across the front.
%When written in components, the boundary conditions are
%\begin{eqnarray}
%\label{eq-boundary1}
%\left[H_{x}\right]+\partial_{x}{Z}_{f}\left[H_{z}\right]=0,\\
%\label{eq-boundary2}
%\left[B_{z}\right]-\partial_{x}{Z}_{f}\left[B_{x}\right]=0.
%\end{eqnarray}
 After resolving the boundary conditions and Maxwell's equations, we find the relations between the field perturbations at the front, at $z=0$, and the front perturbations for any Fourier mode, $\widetilde{B}_{z1}= \widetilde{B}_{z2}= \mu_{0}M k \widetilde{Z}_{f}$,
%\begin{equation}
%\label{eq-linear1}
%\widetilde{B}_{z1}= \widetilde{B}_{z2}= \mu_{0}M k \widetilde{Z}_{f},
%\end{equation}
which reflects the increase of the magnetic field close to the perturbation humps, in agreement with Fig.~\ref{fig:mag_field}. Within the linear stability problem, the perturbations of the front velocity are calculated as
$\partial_{t} \widetilde{Z}_{f} = U'_{f} \widetilde{B}_{z}$, with $U'_{f}\equiv dU_{f}/dB$, and we find the dispersion relation
\begin{equation}
\label{dispersion}
\sigma =k U'_{f} \mu_{0} M.
 \end{equation}
Thus, a thin magnetization-switching front is unconditionally unstable against multidimensional perturbations bending the front.
The structure of the dispersion relation, $\sigma \propto k$, is mathematically similar to the Darrieus-Landau instability of a flame front encountered in combustion, astrophysics, and laser fusion \cite{Law-book,Bychkov-Liberman-2000,Modestov-et-al-2009,DL-supernovae}, and to the instability of doping fronts in organic semiconductors \cite{Bychkov-et-al-2011,Bychkov-et-al-2012}. The similarity of these dispersion relations implies complex multidimensional dynamics of  magnetic fronts, analogous to flames, with the possibility of cellular and fractal structures emerging at the fronts \cite{Law-book,Bychkov-Liberman-2000}. Still, as we show below, the magnetic instability demonstrates also some unique features, such as powerful  front acceleration, which does not happen for the traditional Darrieus-Landau instability.

The characteristic strength of the new instability $\sigma/U_{f}k$
%, as compared to the intrinsic front propagation,
is determined by  the magnetization  $M$ and the sensitivity of the
front speed to the magnetic field perturbations $U'_{f}/U_{f}$. In
particular, in the case  of permalloy nanowires, the so-called
``viscous'' (i.e. controlled by dissipations) regime of domain-wall propagation corresponds to a front
speed proportional to the applied field, $U_{f}\propto H$, with the
proportionality factor about $1.1\,\textrm{m}^{2}/\textrm{sA}$ (see
Ref.~\cite{Nakatani-NatMat-2003}). Taking $\mu_{0}M=1\,\textrm{T}$ for a permalloy and typical domain wall speed of
$U_{f}\sim 500\, \mathrm{m/s}$, we obtain an extremely strong instability with $\sigma/U_{f}k \sim 10^{3}$. In that case even minor front bending modifies the magnetic field strongly, with a considerable increase of the propagation speed of the domain wall.

\begin{figure}
\includegraphics[width=3.4in]{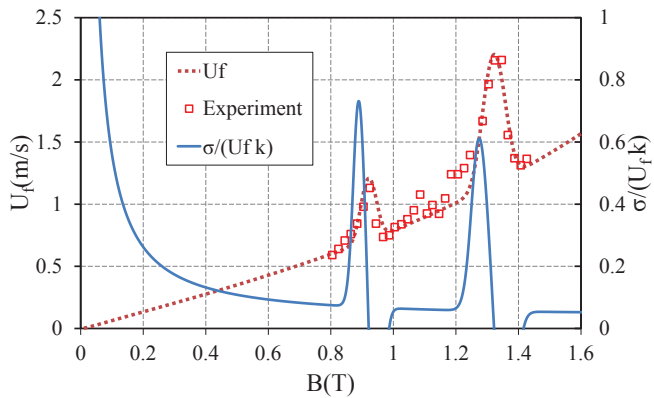}
\caption{\label{fig:growth_rate}The magnetic deflagration speed $U_{f}$ and the scaled instability growth rate, $\sigma / U_{f} k$, vs magnetic field, with $k=2 \pi / \lambda$, for magnetization $\mu_{0}M=0.05 \textrm{T}$. The markers present the experimental data for $U_{f}$ obtained in \cite{Hernandez-PRL-05}. The parameters of the resonance peaks at $B_{1,2}=0.92; 1.32 (\textrm{T})$ are $a_{1,2}=1.89; 2.61$ and $b_{1,2}=840;870$.}
\end{figure}
In contrast to ferromagnetic materials, the magnetization of crystals
of nanomagnets is rather moderate, corresponding to $\mu_{0}M\approx
0.05\,\mathrm{T}$~\cite{Garanin-2012}.  We take the dependence of the
magnetic deflagration speed $U_{f}$ on the applied magnetic field from
the experimental work in Ref.\ \cite{Hernandez-PRL-05}, shown by
markers on Fig.~\ref{fig:growth_rate} and fitted by the red curve.
The curve reflects a monotonic increase of the magnetic
deflagration speed with the field but for local peaks of $U_{f}$ due
to quantum resonances at $B\approx 0.92\,\mathrm{T}$ and
$1.3\,\mathrm{T}$. There are more resonances in the dependence, which have not been measured in Ref.\ \cite{Hernandez-PRL-05}, but may be found, e.g. in
Ref.\ \cite{Subedi-2013}.
The monotonic part of the dependence may be
described by a simple formula originating from combustion theory
\cite{Garanin-Chudnovsky-2007,Modestov-2011,Dion-2013}, as $U_{f} =
\sqrt{\kappa/\tau_{R}\textrm{Ze}}\exp{\left(-{E}_{a}/2T_{f}\right)}$,
where
%$\kappa$ is thermal diffusion,
$\tau_{R}$ is the characteristic
time of spin flipping and $\textrm{Ze}\sim E_{a}/4T_{f}$ is the
Zeldovich number.
The activation energy $E_{a}$ (in temperature units) and temperature
$T_{f}$ behind the front are determined by the applied magnetic field,
see \cite{Garanin-Chudnovsky-2007,Modestov-2011,Dion-2013} for
details.
%The theoretical velocity curve in Fig. 2 has been produced assuming a planar front shape; however, the front is presumably unstable and curved in that domain, and therefore we present the respective plot by the dashed line to indicate the uncertainty.
In the present work we
describe the resonances in $U_{f}$ by taking $\tau_{R}$ as a function
of the magnetic field with the local resonance peaks approximated by the Gaussian  function: $\tau_{R}=\tau_{0}/\left\{1+\Sigma
  a_{i}\exp[-b_{i}(B/B_{i}-1)^2]\right\}$, where $B_{i}$ is the
respective resonance field, and the parameters $a_{i}, b_{i}$ control
the height and width of the resonance. A similar Lorentzian shape of the resonance peaks has been suggested in the analytical model \cite{Michael-00}. We take the resonance width and height
as free parameters of the problem and calculate the relative
instability strength $\sigma/U_{f}k$ in crystals of nanomagnets as
presented in Fig.~\ref{fig:growth_rate}. In the chosen magnetic field
domain, we observe three regions of considerable instability strength
$\sigma/U_{f}k \sim 1$: at low magnetic fields $B < 0.4\, \mathrm{T}$
when the front velocity is small, and close to the resonances, when
the front velocity is sensitive to the field perturbations.

\begin{figure}
\includegraphics[width=3.4in]{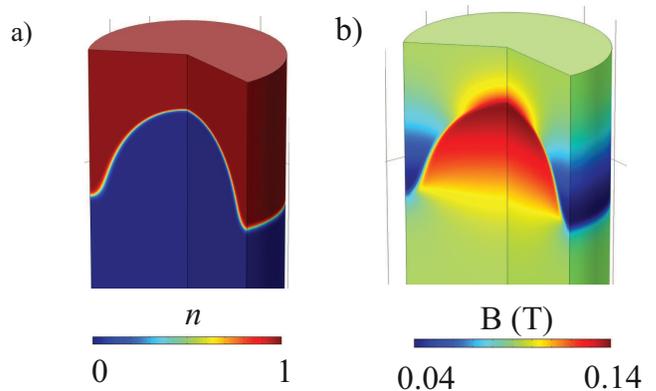}
\caption{\label{fig:curved_front}A stationary curved magnetic
  deflagration front in the axisymmetric geometry obtained numerically
  using Eqs. (4), (5) for the external magnetic field $B_{0}=0.1\,\mathrm{T}$, magnetization $\mu_{0}M=0.05\,\textrm{T}$ and the scaled channel radius $R/L_{f}=21$. (a) Fraction of
  molecules in the metastable state, $n$. (b) Magnetic field
  magnitude.}
\end{figure}

As the  amplitude of front perturbations grows, nonlinear effects
 become important with a possible saturation of
the instability growth to a stationary (i.e. time-independent) cellular structure. We here
solve the nonlinear problem of a stationary cellular front propagating with constant speed by using
the classical Layzer model, which has been employed successfully
within the theory of the Rayleigh-Taylor instability  \cite{Bychkov-Liberman-2000}.
To be particular, in the nonlinear problem we consider an axisymmetric pattern of the curved front, which reproduces the most important quantitative properties of the respective 3D geometry, and still retains quasi-2D simplifications from the analytical and numerical points of view. For comparison, it has been demonstrated that velocity increase of a curved Darrieus-Landau unstable flame is practically independent of a particular 3D or axisymmetric front shape \cite{Bychkov-Liberman-2002}.
Within the Layzer
model, the magnetic field is approximated by the leading Fourier modes
ahead of and behind the front, which are matched at the tip of the
curved stationary front. Specifically, we consider the axisymmetric
cell geometry shown in Fig.~\ref{fig:curved_front} and take the
magnetic field in the form $\textbf{B}_{0}+\widetilde{\textbf{B}}_{1,2}$,
with $\widetilde{\textbf{B}}=-\nabla\phi$, the scalar potential
$\phi=\Phi_{1,2}\exp\left(\mp k z\right)J_{0}\left(kr\right)$ and the
zero-order Bessel function $J_{0}$. The amplitudes $\Phi_{1,2}$ are determined by the boundary conditions at the bent front.
The front shape at the tip is
parabolic, $Z(r)=-\alpha r^2$, where the coefficient $\alpha$ has to
be found from the problem solution. By substituting the obtained
magnetic field into the boundary conditions, we find an increase of
the field at the front tip, $\tilde{B}_{z}(0)\equiv \tilde{B}_{0}= 8 M
\mu_{0} \alpha / k$, so that the front tip propagates at an increased
speed $U_{f}(B_{0}+\tilde{B}_{0})$. Since all points of a stationary
front propagate at the same speed, we arrive at the equation
\begin{equation}
 \label{eq:front_vel}
 U_{f}\left[B_{0}+\widetilde{B}_{0} J_{0}(kr)\exp(kz)\right]= \widehat{\textbf{a}}_{z}\cdot\widehat{\textbf{a}}_{n} U_{f}\left(B_{0}+\widetilde{B}_{0}\right),
 \end{equation}
where $\widehat{\textbf{a}}_{n}$ is a normal vector to the front surface at $Z(r)$. Expanding Eq.~(\ref{eq:front_vel}) at the front tip in $k z \ll 1$, $k r \ll 1$, we obtain
 \begin{equation}
 \label{eq:fild_incr}
 \widetilde{B}_{0}=\frac{8\mu_{0}^{2}M^2 U'_{f}}{U_{f}-4\mu_{0}M U'_{f}},
 \end{equation}
where $U_{f}$ and $U'_{f}$ are taken at $B_{0}+\widetilde{B}_{0}$. The solution to Eq.~(\ref{eq:fild_incr}) determines the increase of the magnetic field at the front tip, $\widetilde{B}_{0}$, and hence the stationary front propagation speed. An important feature of Eq.~(\ref{eq:fild_incr}) is the lack of a stationary solution for a sufficiently strong dependence of the front speed on the magnetic field, $4\mu_{0}M U'_{f}/U_{f}>1$. In that case, a powerful front acceleration with increasing curvature is expected with no saturation, until additional physical effects come into play and limit the front speed.

%\begin{figure}
%\includegraphics[width=3.4in]{fig4.eps}
%\caption{\label{fig:vel_theory} Velocity of the  front vs longitudinal magnetic field.  Dashed line corresponds to the unstable regime of the deflagration.  parameters are $a_{1}=1.9$, $B_{1}=0.92$ T, $b_{1}= 10^{3}T^{-2}$}
%\end{figure}

%After solving Eq.(\ref{eq:fild_incr}) we find field increment as function of external field: $\widetilde{B}_{0}(B_{0})$. Now we can build dependence of the curved front velocity vs external field $U_{f}\left(B_{0}+\widetilde{B}_{0}(B_{0})\right)$ shown in Fig. \ref{fig:vel_theory}. We will take into account only solutions with $\widetilde{B}_{0}>0$, because solution with $\widetilde{B}_{0}<0$ corresponds to the negative curvature of the front. We see that curved front moves faster than the planar one.  Velocity dependence next to the resonance pick has complex nonlinear behavior. Since phenomena of the tunneling resonance is not fully reviled, it gives us  right to variate resonance parameters.  In particular in Fig.\ref{fig6} b) we take arbitrary parameters in order to demonstrate properties of the  phenomena. Area marked by blue is characterized by $dU_{f}/dB_{2}>0$, i,e. there is stable to the front perturbations. Area marked by red is characterized by strong instability, i.e. at such fields front has magnetic field on a tip which corresponds to the resonance maximum $B_{1}$.

We have also validated the nonlinear theory by direct
numerical simulations of the magnetic deflagration fronts for 2D and axisymmetric geometries using the
basic equations of energy transfer and kinetics of spin flipping,
\begin{equation}
\frac{\partial E}{\partial t} = \nabla \cdot (\kappa \nabla E) -  Q
\frac{\partial n}{\partial t}, \label{eq:dEdt}
\end{equation}
\begin{equation}
\frac{\partial n}{\partial t} = - \frac{1}{\tau_R} \exp \left( -
\frac{E_a}{T}\right)  \left[ n - \frac{1}{\exp (Q/T)+1} \right],
\label{eq:dndt}
\end{equation}
where $E$ is thermal (phonon) energy, $n$ is the fraction of nanomagnets in the metastable state,
%which changes from 1 ahead of the front to 0 behind the front in the case of complete spin-flipping,
and $Q$ is the Zeeman energy release determined by the magnetic field
at the front, see Refs.\
\cite{Garanin-Chudnovsky-2007,Modestov-2011,Dion-2013} for
details. Here $E$, $E_{a}$ and $Q$ are taken in temperature units.
Equations (\ref{eq:dEdt}) and (\ref{eq:dndt}) have been complemented by Maxwell's magnetostatic equations.
%We performed the numerical simulations for the 2D and 3D-axisymmetric geometries using finite element COMSOL multiphysics software package.
The initial temperature was taken to be uniform and low, $T_{0}=0.1\,\mathrm{K}$, but for the small region close to the bottom of the computational domain, where it was raised to $T_{f}=30\,\mathrm{K}$ required to induce the spin-flipping process. Slight bending of the hot region initiated the instability development.  Boundaries of the computational domain are thermally insulating with $\hat{\textbf{a}}_{n}\cdot  \nabla T =0$; we also take  $\hat{\textbf{a}}_{n}\cdot \textbf{B} =0$ at the side boundaries and uniform $B_{0}$ at the top/bottom of the domain.
%Constitutive relation is given by (\ref{constitutive relation}). Magnetization of the media is given by formula $M_{z}=(1-2 n)M$, where saturation magnetization was taken $M=0.05$ T.

\begin{figure}
\includegraphics[width=3.4in]{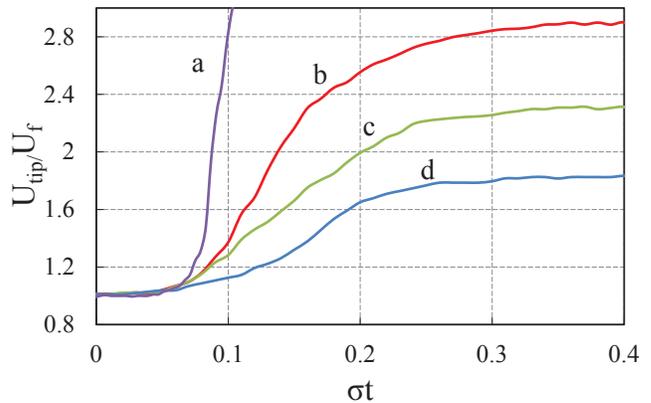}
\caption{\label{fig:vel_dyn}Scaled velocity of the front tip, $U_{\mathrm{tip}}$, versus scaled time for $B_{0}=0.86$ T close to the first resonance for $b_{1}=840 $ and   $a_{1}= 30$ (a);   $a_{1}= 10$ (b);    $a_{1}= 5$ (c);  $a=1.89$ (d).}
%and the scaled channel radius $R/L_{f}=5$. }
\end{figure}

Figures \ref{fig:mag_field} and \ref{fig:curved_front} show the
characteristic shape of the curved stationary fronts
obtained numerically as a result of the instability development for
the 2D and axisymmetric geometries, respectively. Similar to
stationary corrugated flames and doping fronts, the cellular
multidimensional structure of the spin-avalanche fronts may be
described as smooth humps facing the initial cold material and sharp
cusps pointing at the transformed matter behind the front. The
numerical modeling demonstrates also a strong increase of the magnetic
field at the smooth tip, and a decrease of the field at the cusps, in
agreement with the presented theory. In the numerical solution, we
have also reproduced the regime of powerful acceleration for the cases
of strong dependence of the front speed on the magnetic field close to
the quantum resonances. In particular, for the first quantum resonance
field $B_{1}=0.92\,\mathrm{T}$ in Fig.~\ref{fig:growth_rate}, with the
width of the peak set by the parameter $b_{1}=840$ similar
to the experimental data \cite{Hernandez-PRL-05}, the regime of
powerful acceleration takes place for an applied field
$B_{0}>0.84\,\mathrm{T}$. In this regime, the Huygens nonlinear
stabilization of the front bending, which is common for flames
\cite{Bychkov-Liberman-2000}, cannot stop the development of the
instability, and the magnetic deflagration accelerates until the tip
speed reaches the limiting speed characteristic for the quantum
resonance peak. Here we stress that acceleration of this type is a unique feature of the magnetization fronts; the Darrieus-Landau instability in combustion, laser plasma or astrophysics does not exhibit any effect of this kind. Figure~\ref{fig:vel_dyn} shows the acceleration of the
spin avalanche close to the magnetic resonance with the resonance
heights set by the parameter $a_{1}= 1.89; 5; 10; 30$; the value $a_{1}=
1.89$ stems from the experimental data \cite{Hernandez-PRL-05}. At the
same time, the theoretical model \cite{Garanin-2012} of the quantum
resonances suggests an ultimately large resonance height, well above
the sound speed in the crystals, $2000\,\mathrm{m/s}$. Then the
instability may initiate a deflagration-to-detonation transition of
magnetic avalanches, from the strongly subsonic speed of about 1 m/s
to the supersonic speed as observed in the nanomagnet experiments
\cite{Decelle-09}, and similar to the respective combustion process
\cite{Bychkov-et-al-2005,Bychkov-et-al-2008}.

Thus, the experimental signature of the obtained instability is the curved front shape and increased velocity of the magnetization front propagation. 
%At the same time we believe that most of the experimental measurements for spin avalanches in crystals of nanomagnets have been \textit{already} performed for unstable magnetization fronts.
%The optimal way of detecting the instability would be, of course, direct observation of the front shape, which is, unfortunately, hardly possible at present.
One may also expect that the present instability
gives rise to an asymmetric shape of the resonance peaks; still, there
is currently insufficient experimental data to test this expectation.
%For example, development of the instability may explain rather wide and smooth resonance peaks in Fig. 2  as compared to the theoretical model of the resonances \cite{Garanin-2012}.
 %As an alternative way to elucidate the instability, experimental data for the front propagation speed may be compared for the magnetization front propagating along the external magnetic field (when the front is unstable) and perpendicular to the magnetic field (when the front is stable) with other parameters kept the same; experiments of this type are possible but have not been performed so far. To clarify the possibility, we point out that the propagation direction  of the spin avalanches is determined by the temperature gradient, which is controlled by the conditions of the front initiation, e.g., by external heating. Hence, the direction of front propagation may be different from the direction of the easy axis of the crystal and the direction of the magnetic field, see Ref. \cite{Dion-2013} for the detailed explanation of the anisotropic effects of the magnetic deflagration dynamics.
Besides, the instability obtained in the present work may be
responsible for the magnetic deflagration-to-detonation transition
observed experimentally in Ref. \cite{Decelle-09}; the process of magnetic detonation triggering requires more studies.

To summarize, we have obtained a universal multidimensional instability
of magnetization-switching  fronts, which may develop spontaneously in
super-paramagnetic and ferromagnetic media such as crystals of
nanomagnets, ferromagnetic nanowires and systems of quantum dots. The
instability  leads to a curved front structure with a possible strong increase of the propagation speed, and hence allows control of the front dynamics. Due to the universal instability properties, we expect our results to be applicable to a wide variety of problems.

%\bigskip

%\textbf{Acknowledgments}

%\textbf{This work was supported by the Swedish Research Council and by the Kempe Foundation. The authors thank Petter Minnhagen, Bertil Sundqvist, Thomas W{\aa}gberg, Sune Pettersson, Tatiana Makarova and Valeria Zagainova for useful discussions.}


\begin{thebibliography}
\bibliographystyle{}

\bibitem{Bogani-Wernsdorfer-2008}
L. Bogani, W. Wernsdorfer,
Nature Mater. \textbf{7}, 179 (2008).

%\bibitem{Gimenez-Lopez-et-al-2011}
%M. Gimenez-Lopez, F. Moro, A. La Torre, C. J. Gomez-Garcia, P. D. Brown, J. van Slageren, A. N. Khlobystov
%Nature Commun. \textbf{2}, 407 (2011).

\bibitem{Nakatani-NatMat-2003}
Y. Nakatani, A. Thiaville, and J. Miltat, Nature Mater. \textbf{2}, 521 (2003).

\bibitem{Beach-NatMat-2005}
G. S. D. Beach, C. Nistor, C. Knutson, M. Tsoi, and J. L. Erskine,
Nature Mater. \textbf{4}, 741 (2005).

\bibitem{Misiorny-NatPhys-2013}
M. Misiorny, M. Hell, and M. R. Wegewijs, Nature Phys. \textbf{9}, 801 (2013).

%\bibitem{Gatteschi-review-03}
%D. Gatteschi and R. Sessoli, Angew. Chem., Int. Ed. \textbf{42}, 268 (2003).

%\bibitem{Barco-review-05}
%E. del Barco, A. D. Kent, S. Hill, J. M. North, N. S. Dalal, E. Rumberger, D. N. Hendrikson, N. Chakov, and G Christou, J. Low Temp. Phys. \textbf{140}, 119 (2005).

%\bibitem{Leuenberger-01}
%M. N. Leuenberger and D. Loss, Nature (London) \textbf{410}, 789 (2001).
%\bibitem {Tejada-01}
%J. Tejada, E. M. Chudnovsky, E. del Barco, J. M. Hernandez, and T. P.
%Spiller, Nanotechnology \textbf{12}, 181 (2001).

%\bibitem{Sessoli-Nature-93}
%R. Sessoli, D. Gatteschi, A. Caneschi, and M. A. Novak, Nature (London) \textbf{365}, 141 (1993).

\bibitem {Friedman-PRL-96}
J. R. Friedman, M. P. Sarachik, J. Tejada, and R. Ziolo, Phys Rev. Lett.
\textbf{76}, 3830 (1996).

\bibitem{Thomas-Nature-96}
L. Thomas, F. Lionti, R. Ballou, D. Gatteschi, R. Sessoli, and B. Barbara,
Nature (London) \textbf{383}, 145 (1996).

%\bibitem{Forminaya-97}
%F. Fominaya, J. Villain, P. Gandit, J. Chaussy, and A. Caneschi, Phys. Rev.
%Lett. \textbf{79}, 1126 (1997).
%\bibitem{Barco-99}
%E. del Barco, J. M. Hernandez, M. Sales, J. Tejada, H. Rakoto, J. M. Broto,
%and E. M. Chudnovsky, Phys. Rev. B \textbf{60}, 11898 (1999).

%\bibitem{McHugh-07}
%S. McHugh, R. Jaafar, M. P. Sarachik, Y. Myasoedov, A. Finkler, H.Shtrikman, E. Zeldov, R. Bagai, and G. Christou, Phys. Rev. B \textbf{76}, 172410 (2007).

%\bibitem{Hernandez-08}
%A. Hernandez-Minguez, F. Macia, J. M. Hernandez, J. Tejada, and P. V. Santos, J. Magn. Magn. Mater. \textbf{320}, 1457 (2008).

\bibitem{Suzuki-05}
Y. Suzuki, M. P. Sarachik, E. M. Chudnovsky, S. McHugh, R. Gonzalez-Rubio,
N. Avraham, Y. Myasoedov, E. Zeldov, H. Shtrikman, N. E. Chakov, and G.
Christou, Phys. Rev. Lett. \textbf{95}, 147201 (2005).

\bibitem{Hernandez-PRL-05}
A. Hern\'{a}ndez-M\'{\i}nguez, J. M. Hernandez, F. Maci\`{a},
A. Garc\'{\i}a-Santiago, J. Tejada, and P. V. Santos,
Phys. Rev. Lett. \textbf{95}, 217205 (2005).

\bibitem{Garanin-Chudnovsky-2007}
D. A. Garanin and E. M. Chudnovsky, Phys. Rev. B. \textbf{76}, 054410
(2007).

\bibitem{Modestov-2011}
M. Modestov, V. Bychkov, and M. Marklund, Phys. Rev. B \textbf{83} 214417 (2011).


%\bibitem{Villuendas-08}
%D. Villuendas, D. Gheorghe, A. Hernandez-Minguez, F. Macia, J. M. Hernandez, J. Tejada, R. J. Wijngaarden, EPL (Europhysics Letters) \textbf{84}, 67010 (2008).

\bibitem{McHugh-09}
S. McHugh, B. Wen, X. Ma, M. P. Sarachik, Y. Myasoedov, E. Zeldov, R. Bagai,
and G. Christou, Phys. Rev. B \textbf{79}, 174413 (2009).

\bibitem{Subedi-2013} P. Subedi, S. V\'{e}lez, F. Maci\`{a}, S. Li,
  M. P. Sarachik, J. Tejada, S. Mukherjee, G. Christou, and
  A. D. Kent, Phys. Rev. Lett. \textbf{110}, 207203 (2013).

\bibitem{Dion-2013}
C. M. Dion, O. Jukimenko, M. Modestov, M. Marklund, and V. Bychkov, Phys. Rev. B \textbf{87} 014409 (2013).


\bibitem{Decelle-09}
W. Decelle, J. Vanacken, V. V. Moshchalkov, J. Tejada, J. M.
Hern\'{a}ndez, and F. Maci\`{a}, Phys. Rev. Lett. \textbf{102}, 027203 (2009).


\bibitem{Modestov-det}
M. Modestov, V. Bychkov, and M. Marklund, Phys. Rev. Lett. \textbf{107}, 207208 (2011).

%\bibitem{Jaafar-et-al-2008}
% R. Jaafar, S. McHugh, Y. Suzuki, M. P. Sarachik, Y. Myasoedov, E. Zeldov, H. Shtrikman,  R. Bagai, and G. Christou, J. Magn. Magn. Mater. \textbf{320}, 695 (2008).

\bibitem{Garanin-2013}
D. A. Garanin, Phys. Rev. B \textbf{88} 064413 (2013).

%\bibitem{LL-Fluidmechanics}
%L. Landau and E. Lifshitz, \textit{Fluid Mechanics}, Pergamon Press, Oxford, 1989.

\bibitem{Law-book}
C. K. Law, \emph{Combustion Physics} (Cambridge University Press,
Cambridge, 2006).

\bibitem{Bychkov-Liberman-2000}
V. Bychkov and M. Liberman, Phys. Rep. \textbf{325}, 115 (2000).

\bibitem{Modestov-et-al-2009}
M. Modestov, V. Bychkov, D. Valiev, and M. Marklund, Phys. Rev. E \textbf{80}, 046403 (2009).

\bibitem{DL-supernovae}
J. B. Bell, M. S. Day, C. A. Rendleman, S. E. Woosley, and M. Zingale,
Astrophys. J. \textbf{606}, 1029 (2004).

\bibitem{Bychkov-et-al-2011} V. Bychkov, P. Matyba, V. Akkerman,
  M. Modestov, D. Valiev, G. Brodin, C.K. Law, M. Marklund, and
  L. Edman, Phys. Rev. Lett. \textbf{107}, 016103 (2011).

\bibitem{Bychkov-et-al-2012} V. Bychkov, O. Jukimenko, M. Modestov,
  and M. Marklund, Phys. Rev. B \textbf{85}, 245212 (2012).

\bibitem{Garanin-2012}
D. A. Garanin and S. Shoyeb, Phys. Rev. B \textbf{85}, 094403 (2012).

%\bibitem{Dorofeev-2011}
%S. Dorofeev, Proc. Combust. Inst. \textbf{33} 2161 (2011).

\bibitem{Michael-00}
M. N. Leuenberger and D. Loss, Phys. Rev. B \textbf{61}, 1286.

\bibitem{Bychkov-Liberman-2002}
V. Bychkov, M. Liberman, Phys. Fluids \textbf{14},
2024 (2002).

\bibitem{Bychkov-et-al-2005}
V. Bychkov, A. Petchenko, V. Akkerman, and L.-E. Eriksson, Phys. Rev. E
\textbf{72}, 046307 (2005).


\bibitem{Bychkov-et-al-2008}
V. Bychkov, D. Valiev, and L.-E. Eriksson, Phys. Rev. Lett. \textbf{101}, 164501
(2008).

%\bibitem{Valiev-et-al-2010}
%D. Valiev, V. Bychkov, V. Akkerman, C. K. Law, L.-E. Eriksson, Combust. Flame \textbf{157}, 1012 (2010)

%\bibitem{Prokof'ev-97}
%N.V. Prokof'ev and P.C.E. Stamp, Phys. Rev. Lett. \textbf{80}, 5794 (1998).



\end{thebibliography}
\end{document}